\documentclass[referee,useAMS,usenatbib]{mn2e}
\usepackage{graphicx}

\title[Three low surface brightness dwarfs discovered around NGC 4631]{Three low surface brightness dwarfs discovered around NGC 4631}

 \author[I.D.Karachentsev et al.]{Igor D. Karachentsev$^{1}$,  Dirk Bautzmann$^{2}$,
 Fabian Neyer$^{2}$, Robert P\"{o}lzl$^{2}$, \and Peter Riepe$^{2}$, Thorsten Zilch$^{2}$, and Bruno Mattern$^{2}$\\
$^{1}$Special Astrophysical Observatory of the Russian Academy of Sciences\\
$^{2}$TBG group in the German amateur association\thanks
{TBG is a group in the German amateur association ``Vereinigung der
Sternfreunde e.V. '' called ``Tief Belichtete Galaxien'',
which performs the project ``Very long exposed galaxies''.}}
\begin{document}

\date{Accepted December 2013}

\pagerange{\pageref{firstpage}--\pageref{lastpage}} \pubyear{2013} \onecolumn

\maketitle

\label{firstpage}

\begin{abstract}

 We report the discovery of three low surface brightness companions to the
spiral galaxy NGC~4631, made with small amateur telescopes. Assuming their
distances to be 7.4 Mpc, the same as that of NGC~4631, the absolute magnitudes
and linear diameters of the dwarfs are ranged within [-12.5, -9.6] mag and
[4.7 - 1.3] kpc, respectively. These new three dwarfs, together with the
discovered by us diffuse structure called  ``bridge'', look like parts of
a tidal filament directed towards NGC~4656 at total extended over 100 kpc.
 \end{abstract}


 \section{Introduction}
 Wide-field surveys of the halo of the Andromeda galaxy performed with
3-meter class telescopes designed with the CCD-mosaics led to the discoveries of many
ultra-faint dwarf companions and low surface brightness (LSB) stellar streams \citep{b7,b12,b5}.
A similar survey
of wide surroundings of another nearby giant spiral galaxy M 81 \citep{b3,b2}
also demonstrated the presence around it of many
dwarf satellites of very low surface brightness, SB$\sim(26-28)m/\sq\arcsec$.
Diffuse objects having a surface brightness of $\sim1$\% above the moonless
night sky are usually indistinguishable on the photographic plates of sky surveys
POSS-I, POSS-II, ESO-SERC and reveal themselves only being resolved into stars.

The brightness limit of SB$\sim27m/\sq\arcsec$ is easily reachable for the
telescopes designed with modern CCD mosaics under a long enough exposure time
and accurate flat-fielding. Because the surface brightness of a galaxy does
not depend on its distance, a search for new very LSB objects can be successfully
performed with a telescope of small diameter ($\sim0.1-1$ meter) having a wide
field of view. For instance, a deep image of a galaxy situated at a distance
of 10 Mpc, carried out with a $\sim1$ Mega-pixel camera under a resolution of
$\sim3\arcsec$/pixel, covers with its field of view $\sim1^{\circ}$ the sky region of
$\sim100$ kpc, comparable with the size of the halo of a giant galaxy. Hereby, the image
of a dwarf companion having a typical diameter of $\sim1$ kpc (or $20\arcsec$)
contains a number of resolution elements ($\sim50$), which is sufficient to
reliably detect the object.

 According to the standard cosmological paradigm, luminous galaxies are building up
from dwarf galaxies via their consecutive merging. Due to this, the far
periphery of nearby giant galaxies looks like a screen where one can see
stellar streams as a result of dwarf galaxy merging, and a population of LSB
dwarfs missed by the photographic sky surveys.

  These apparent considerations were set as a ground for different projects
which directed astronomy amateurs, having small-gauge telescopes, to perform
a systematic survey of the vicinities of nearby luminous galaxies. The efficiency
of such a strategy was demonstrated by \citep{b13,b14,b15}
whose project ``Dwarf Galaxy Survey with Amateur Telescopes = DGSAT'' led to the
discovery and study of tidal stellar streams around several nearby spiral
galaxies.

 A similar project, called TBG (Tief Belichtete Galaxien) was started by
the German amateur association ``Vereinigung der Sternfreunde e.V.'', which
has several workgroups, the ``Fachgruppe Astrofotografie'' being one of
them. In 2011 this group started a project in collaboration with
professional astronomers from the Bochum University. The project is
called TBG, which stands for the German expression "Tief Belichtete Galaxien"
(meaning ``very long exposed galaxies''). All the selected targets are the galaxies from
the Local Volume. Initially, the project's aim was to detect  new
stellar streams. As shown by the first results, the wide-field photography made by the
team also allows to reveal other interesting details:  the tidal interaction
phenomena between galaxies, as well as the dwarf galaxies of very low surface
brightness. Meanwhile, the TBG team reached SB of $\sim27m/\sq\arcsec$,
which is deeper than that of the Sloan Digital Sky Survey = SDSS \citep{b1}.
 In autumn 2013 the TBG aims were redefined and the cooperation
with professional astronomers was also exchanged. In terms of the public outreach,
there will soon be  a TBG website for the presentation of the team results.
Today the group of amateur astrophotographers counts about
30 members. The telescopes in use range from the small refractors (4 inch)
to the reflectors with apertures up to 44 inches. To get very deep and high quality
image data, high-end off-the-shelf CCD cameras of well-known manufacturers are
used. All the images are carefully calibrated by dark frame subtraction and flat
field correction before the further processing is applied. The framework of the
TBG group activities is conceived by Peter Riepe (conceptional work) and
Thorsten Zilch (organization).

 \section{The group of galaxies around NGC~4631}

According to \citet{b11},
a group of galaxies around
NGC~4631 accounts 28 members with a wide range of absolute magnitudes $M_B$
from $-20^m$ till $-10^m$. The group is characterized by the mean radial
velocity of +635 km/s in the rest frame of the Local Group, the velocity
dispersion of 90 km/s, and the mean harmonic radius of 243 kpc. The total
stellar mass of the group members amounts to $\log (M_*/M_{\odot}$) = 11.12,
and the virial (projected) mass of the group is equal to $\log(M_p/M_{\odot})
=12.98$, what is 72 times the stellar mass. The brightest group member
is a spiral late-type galaxy NGC~4631 seen nearly edge-on. On its northern
side, an elliptical galaxy NGC~4627 is projected, and on the southern side
at the separation of 32$\arcmin$ there is another bright late-type spiral
galaxy NGC~4656 showing strong distortions of the disk shape. Radburg-Smith
et al. (2011) determined the distance to NGC~4631 as 7.38 Mpc, via the Tip
of Red Giant Branch using the ACS camera aboard the Hubble Space Telescope.
With this distance, NGC~4631 has a linear Holmberg diameter of 33.7 kpc
and an absolute magnitude of $M_B=-20\fm28$.

  The moderately wide interacting pair NGC~4631/4656 with the linear projected
separation of 69 kpc was a subject of investigations by various authors.
Rand (1994) performed a 21-cm survey of the pair using the WSRT radio telescope
with a resolution of $\sim30\arcsec$ and found four wide HI-tails. Two of them
extend from NGC~4631 towards NGC~4656 and two others are directed to the north and
north-east from NGC~4631. Apart from them,  a ring-like
structure consisting of low-density HI-clouds was detected. These features look like
a tidal bridge with tails created during a tight encounter of two  spiral
galaxies. Judging from their projected separation and radial velocity difference
of 55 km/s, this event happened $\sim1$ Gyr ago.

\citet{b4} imaged the NGC~4631/4656 pair with a narrow
($\sim30$\AA) filter centered on the Balmer $H\alpha$ line. They found
a broad field of faint $H\alpha$ emission, occupying a space between
the pair components and also extending north from NGC~4631. The discovery
of a faint emission envelope around the pair has again confirmed  the presence
of rudimental tidal features remaining from the encounter.

   \section{Discovery of three new dwarf galaxies near NGC~4631}

  To test the own capabilities, the TBG members concentrated their initial
photographic efforts on the already known objects, i.e. the objects imaged
by the group of David Martinez-Delgado and his amateur astronomers \citep{b13}.
As the second step they decided to establish
a new list of targets from the NGC/IC catalog by a  systematic examination
of potential candidates using the Deep Sky Survey (DSS). To identify the
possible candidates, a heavy contrast enhancement for all the images
was applied. Applying this method, some dozens of galaxies, showing
suspicious features like possible stellar streams were selected.
Due to the limited magnitude of the DSS, the target candidates were
compared with the SDSS images, whenever available. The galaxy NGC~4631
turns out to be among the first targets in the project.

\begin{figure*}
 \includegraphics[scale=0.3]{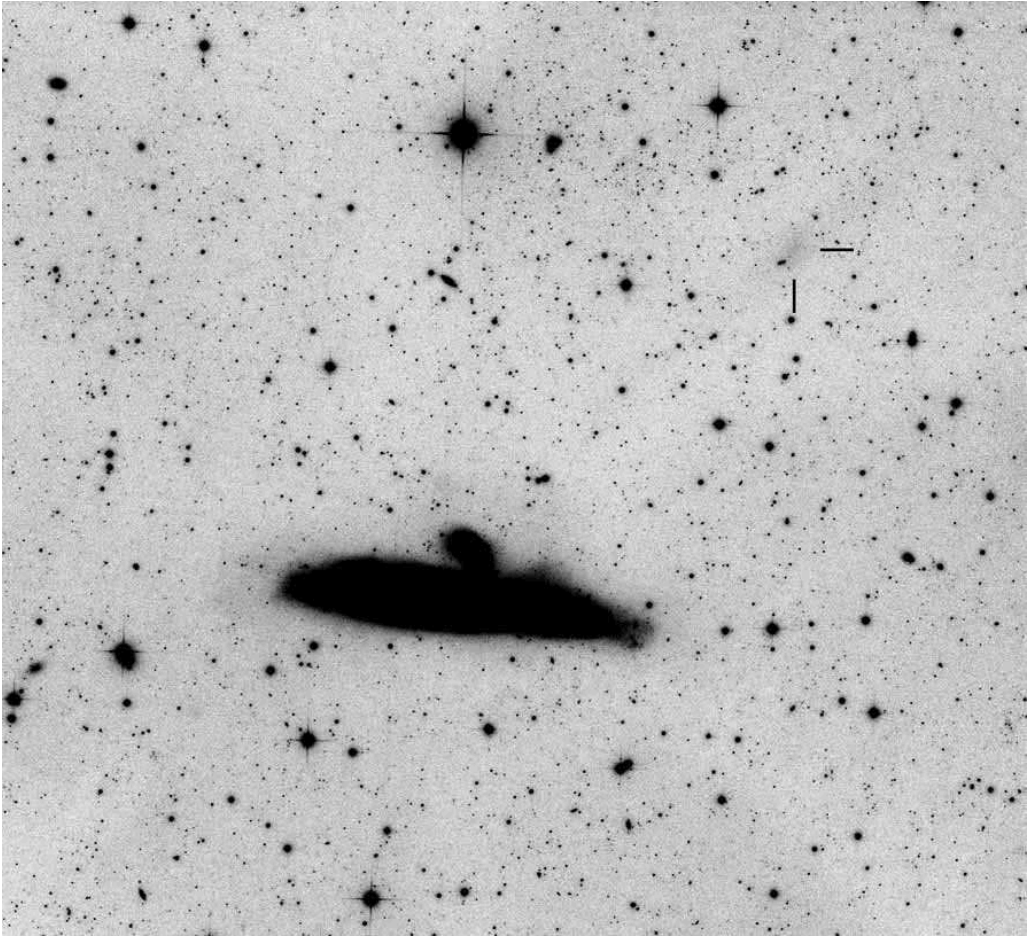}
 \caption{NGC~4631 and N4631dw1 (arrows) taken in March 2013 by
Dirk Bautzmann. Telescope: PlaneWave CDK 12.5 (f/8) on the  ASA DDM85 mount,
camera: Apogee U16M with LRGB+$H\alpha$ filters. Total exposure: 13 h 05 min,
FWHM about 2$\farcs$6, field of view: 41$\arcmin\times 37\arcmin$.}
\end{figure*}

\begin{figure*}
 \includegraphics[scale=0.3]{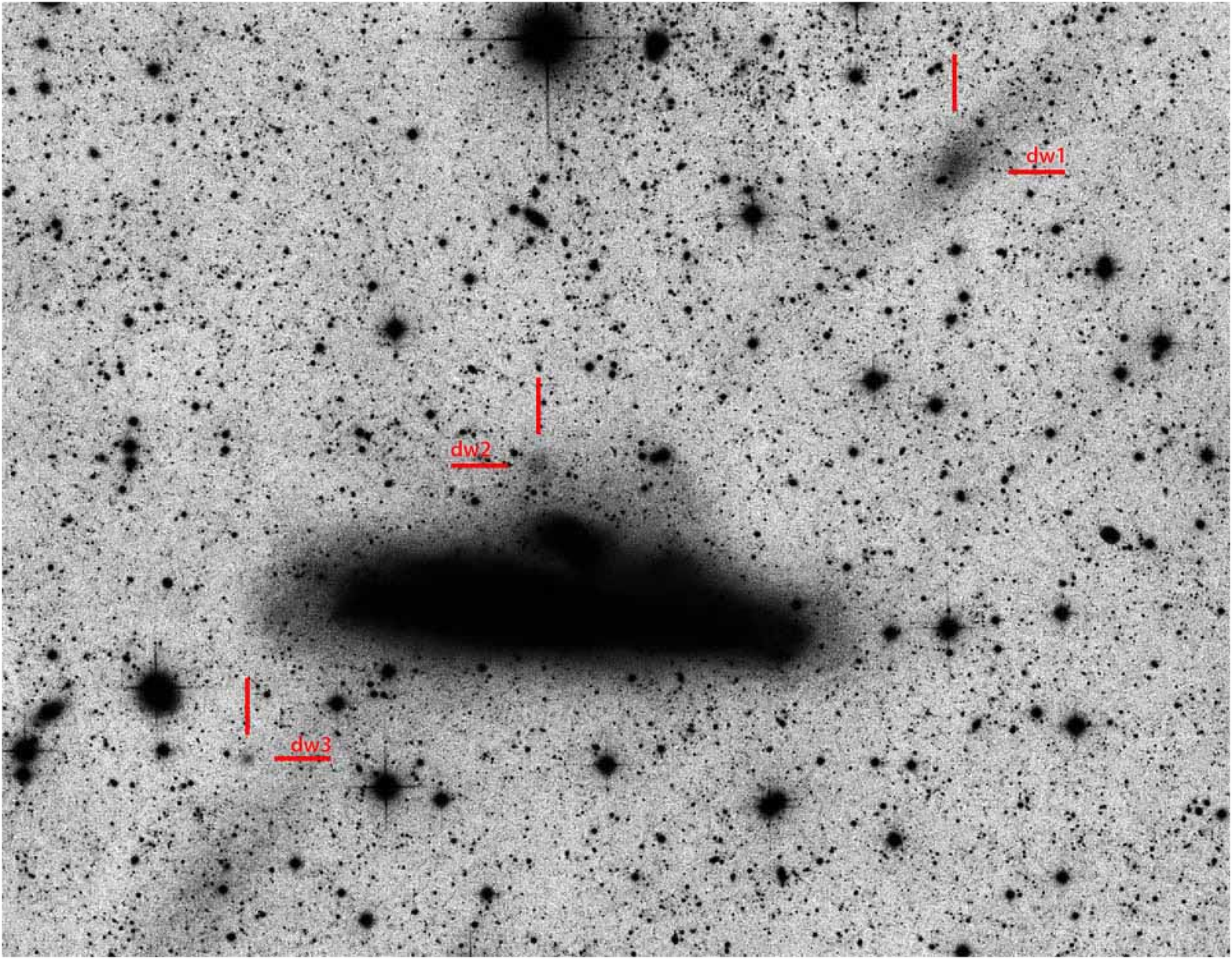}
 \caption{The stacked image of NGC~4631 and three neighboring LSB dwarfs
derived from 75 individual images with a total exposure time of 23.9 hours.
The data consists of: 3.25 hours (only L filter) taken from Dirk Bautzmann in March 2013 with
PlaneWave CDK 12.5 (f/8) on ASA DDM85 mount and Apogee U16M camera +
5.67 hours taken from Robert P\"{o}lzl in May and June 2013, with 14.5
Newton f/3.8 on ASA DDM85 mount and FLI ML 8300 camera + 15 hours taken from
Fabian Neyer in February and March 2013 with 5.5 TEC APO refractor (f/7.2)
on AP900GoTo mount and STL-11000M camera. The average FWHM is 3.0 arcsec,
resampled at 1.80$\arcsec$/pixel, FOV of about 42$\arcmin\times 32\arcmin$, North is up. Processing
by Fabian Neyer. From upper right to lower left the objects: N4631dw1,
N4631dw2, and N4631dw3 are marked. South of the object N4631dw3, the
the mentioned LSB ``bridge'' directed to the NGC 4656 can be seen.}
\end{figure*}

  At first, a wide-field image of the galaxy was derived by Dirk
Bautzmann. A  diffuse  feature 20' northwest from NGC~4631 unknown before
was revealed. It is marked by the bars in Figure 1. Then, Bruno Mattern
from Hamburg identified this feature on his deep wide-field frame
obtained earlier. Later, Robert P\"{o}lzl and Fabian Neyer took follow-up
images of this region. The stacked image of them is presented in Figure 2.
The diffuse object was confirmed again. We called it  N4631dw1.
About 5' north of the NGC~4631 core and 3' north-north-east
of NGC~4627, another diffuse object was found, smaller in size
and likely an extremely faint dwarf galaxy, called N4631dw2. About
10' south-east of the NGC~4631 centre, yet a third faint object is visible,
N4631dw3, looking like a dwarf spheroidal galaxy. At least, on the
opposite side of NGC~4631, compared to N4631dw1, a larger structure
of extremely low surface brightness can be recognized just south of
the  N4631dw3 object. This elongated feature seems to point in the
direction of NGC~4656,  situated in the zone of the HI and $H\alpha$
bridge, found by  \citet{b18} and  \citet{b4}.

 \begin{table*}
    \caption{ Properties of three low surface brightness dwarfs
          discovered around NGC~4631}
  \begin{tabular}{lccccrcl} \\ \hline

\hline
Object   & RA  (J2000.0) DEC  & $a\arcmin$ & $b/a$ & $B_T$& $M_B$ & $m_{FUV}$ &Type\\
\hline
N4631dw1  &  12 40 57.0 +32 47 33  &  2.2 & 0.60 & 16.8 & $-$12.5 & 22.00 & Ir \\
N4631dw2  &  12 42 06.8 +32 37 15  & 0.9  &0.90  &18.5  &$-$10.8  & 22.64  &Ir \\
N4631dw3  &  12 42 52.5 +32 27 35  & 0.6  &0.85  &19.7  &$-$9.6   & ---   & Sph\\
\hline
\end{tabular}
   \end{table*}

  Table 1 presents some basic properties of the  objects, assuming
their apparent membership in the NGC~4631 group: (1) --- the
object name, (2) --- equatorial coordinates for the epoch (J2000.0),
(3) --- angular diameter in arcmin, (4) --- axial ratio,
(5) --- total blue magnitude as estimated by eye-ball, (6) --- absolute
magnitude, (7) --- magnitude in the far ultraviolet band acquired by the GALEX
(http://galex/stsci.edu/GalexView/), which is useful under the morphological
classification of the objects, (8) --- morphological type of the
dwarf: Ir --- irregular, Sph --- spheroidal.

  As follows from these data, all three new dwarf galaxies have
luminosities typical for the  dwarfs of the Local Group and other nearby
groups. Under the scale of $1\arcmin = 2.15$ kpc, linear diameters of the
new objects are also similar to the dwarf population of  the nearby groups.
We do not list the extremely LSB structure, called the ''bridge`` among the
dwarfs as  its large linear dimensions, $16\times 7$ kpc, are unusual
for the known dwarf systems.

   \section{Concluding remarks}

As was noted by \citep{b10,b6,b16},
 a significant part of dwarf satellites around
the Andromeda galaxy and the Milky Way are situated in flat disk-like
structures. The origin of these planar structures is usually explained
as a result of tidal interactions between the luminous galaxies, whose tails
and bridges are transformed after the fragmentation into ordinary dwarfs. In the
case of an interacting pair NGC~4631/4656, three  LSB dwarf galaxies discovered,
along with the diffuse  ''bridge`` structure form a filament (or a
planar system seen edge-on) that extends over 100 kpc. This is
just one more example illustrating the idea that ''tidal dwarfs`` may
be making up a noticeable fraction in the dwarf galaxy population. Herein, \citet{b10}
stress the following piquancy:  tidal dwarfs
principally differ  from the ordinary (relic) dwarfs, because the luminous
matter, rather than the dark matter is their dominating component.

  Considering suites of dwarf galaxies around the nearby luminous galaxies, \citep{b8}
noticed that the number of dwarf companions
tightly correlates with the luminosity (mass) of the principal galaxy.
Among the objects of the "Updated Nearby Galaxy Catalog"  by \citet{b9},
 most of the galaxies with absolute magnitude brighter than $-18.5^m$
possess dwarf companions. There are about 100 bright enough galaxies in
this catalog, whose surroundings are still poorly investigated.
A systematic search of low-contrast signs of interaction and LSB
companions around them using the deep exposures at the amateur telescopes is
an important and actual task of the extragalactic astronomy.

   {\bf Acknowledgments.} This work was supported by the RFBR grants
12--02--90407 and 13--02--92690.

{}

   \end{document}